\documentclass[twocolumn,apl,amsmath,amssymb,showpacs]{revtex4-2}
\usepackage{epsf}      
\usepackage{graphicx}
\usepackage{color}
\usepackage{soul}
\usepackage{gensymb}
\usepackage{sidecap}
\usepackage{amsmath}
\usepackage{mathtools}
\usepackage[hidelinks,colorlinks=true,linkcolor=blue,citecolor=blue]{hyperref}

\begin{document}
\title{Understanding the electrochemical performance and diffusion kinetics of HC$||$Na$_3$V$_2$(PO$_4$)$_3$/C full cell battery for energy storage applications}
\author{Madhav Sharma}
\affiliation{Department of Physics, Indian Institute of Technology Delhi, Hauz Khas, New Delhi-110016, India}
\author{Rajendra S. Dhaka}
\email{rsdhaka@physics.iitd.ac.in}
\affiliation{Department of Physics, Indian Institute of Technology Delhi, Hauz Khas, New Delhi-110016, India}

\date{\today}      

\begin{abstract}
The efficient energy storage devices are crucial to meet the soaring global energy demand for sustainable future. Recently, the sodium-ion batteries (SIBs) have emerged as one of the excellent cost effective solution due to the uniform geographical distribution and abundance of sodium. Here, we use hard carbon (HC) as an anode and Na$_3$V$_2$(PO$_4$)$_3$/C (NVP/C) as a cathode to fabricate a HC$||$NVP/C full cell battery and understand its electrochemical performance and diffusion kinetics. These materials are characterized through the analysis of x-ray diffraction and Raman spectroscopy to confirm their single phase and structure. The full cell demonstrates a high operating voltage of $\sim$3.3 V, with minimal polarization of 0.05 V, attributed to the lower working voltage of the HC. Interestingly, for the full cell battery we find the specific capacity of around 70 mAh/g at 0.1 C and even around 35 mAh/g at high current rate of 5 C along with high rate capability up to 55 cycles. The diffusion kinetics of the full cell battery is investigated through detailed analysis of CV curves at various scan rates, and the diffusion coefficient is found to be 5--8$\times$10$^{-11}$ cm$^2$/s for the anodic as well as cathodic peaks. 

\end{abstract}

\maketitle
\section{\noindent~Introduction}
Sodium-ion batteries (SIBs) have emerged as potential candidates for cost effective energy storage devices to meet the requirements of future generation \cite{Chen_ASS_18, Slater_AFM_13}. Although sodium's larger size and higher standard electrode potential limit the energy density of SIBs compared to lithium-ion batteries (LIBs), the focus on cathode materials with similar intercalation chemistry has facilitated the exploration of viable compounds for both systems \cite{Chen_ASS_18, Slater_AFM_13}. Despite many challenges, ongoing research seeks to optimize SIB performance, making them a compelling candidate for large-scale energy storage applications \cite{Delmas_AEM_18}. A vast category of anodes and cathodes have been discovered, showing potential for their utilization in SIBs, but each has its own merits and demerits. Anodes pose a challenge, with graphite, common in LIBs, showing unsatisfactory sodium ion intercalation \cite{Moriwake_RSCA_17}. However, non-graphitic anodes, particularly hard carbon (HC) derived from various precursors, have emerged as promising alternatives for SIBs due to its remarkable sodium retention ability \cite{Saurel_AEM_18, Dou_MT_19, Xie_PE_20}. Also, HC is an eco-friendly anode option derived largely from biomass and the disordered nature of HCs, enriched with defects as well as hetero-atoms, offers more avenues for Na$^+$ ion storage and diffusion. Moreover, compared to graphite, its expanded interlayer space ensures a stable intercalation/de-intercalation mechanism for sodium ions during charging-discharging process \cite{Saurel_AEM_18, Mittal_JPE_22}. 

For the cathode side, the polyanionic-based materials have emerged as a prominent category for SIBs \cite{Sapra_WEE_21} with Na$_3$V$_2$(PO$_4$)$_3$ (NVP) standing out as a leading candidate due to its noteworthy energy storage capacity and remarkable structural resilience \cite{Jin_CSR_20, Saravanan_AEM_13}. This material boasts a NASICON-type crystal framework characterized by its three-dimensional open lattice that ensures high ionic conductivity \cite{Chen_AM_17}. Nevertheless, a significant challenge encountered with NVP lies in its low electron conductivity; therefore, various strategies have been used to surmount this hurdle, notably incorporating carbon coatings \cite{Wang_SM_19}. There have been many studies on these electrode materials in half-cell configuration; however, detailed investigation of the electrochemical performance in the full-cell configuration are still limited. Here, the selection of both anode and cathode materials plays a critical role in determining the working voltage and energy density of the assembled full cell. A symmetric full cell using NVP has been investigated due to its dual functionality as both a cathode (V$^{3+}$/V$^{4+}$ at $\sim$3.4 V) and an anode (V$^{2+}$/V$^{3+}$ at $\sim$1.7 V) electrode \cite{Ling_CEJ_18}. However, the relatively high potential on the anodic side and a low capacity of around 59 mAh/g lead to a moderate working potential of 1.7~V and low gravimetric energy density in the NVP symmetric full cell \cite{Liu_AMI_19}. While efforts have been made to boost the gravimetric energy density by incorporating high-capacity anode materials in sodium-ion batteries, but giving a nominal voltage of around 2 V only \cite{Panda_NE_19, Chen_MTC_23}, hence there is still room for improvement in achieving an overall higher working potential. To tackle these challenges, the HC is advantageous due to its lower operating potential and higher Na$^+$ ion storage capacity \cite{Zhan_EA_18}.

Therefore, we present a comprehensive analysis of the electrochemical properties of full cell sodium-ion batteries utilizing NVP/C as a cathode and HC as an anode. Using the x-ray diffraction (XRD) and Raman analysis, we confirm the NASICON-type crystal structure of the synthesized cathode material NVP/C. Also, the anode material (hard carbon) displayed characteristics typical of disordered graphite domains. Our electrochemical testing involve cyclic voltammetry (CV), electrochemical impedance spectroscopy (EIS), and galvanostatic charge-discharge (GCD) measurements, which provided insightful outcomes for both NVP/C and HC half cells, as well as the assembled full cell (HC$||$NVP/C). The half cells demonstrated a decent performance, revealing a reversible discharge capacity of 85 mAh/g (operating voltage: $\sim$3.4 V) at 0.1 C for the NVP/C cathode and a charge capacity of 252 mAh/g (operating voltage: $\sim$0.19 V) at 50 mA/g for the HC anode. Interestingly, we get good performance of HC$||$NVP/C full cell having the capacity of around 70 mAh/g at 0.1 C and around 35 mAh/g at 5 C. Additionally, owing to the lower working voltage of HC, the full cell exhibits a notable high operating voltage of $\sim$3.3 V, coupled with minimal polarization of 0.05 V. We perform detailed analysis of CV data at different scan rates to understand the diffusion kinetics of the full cell where the extracted values of the diffusion coefficient are in the range of 5--8$\times$10$^{-11}$ cm$^2$/s for the anodic/cathodic peaks.

\section{\noindent ~Experimental}

{\bf{Synthesis of cathode material:}}

The NVP/C composite sample is synthesized using the sol-gel method where NaH$_2$PO$_4$, V$_2$O$_5$, Na$_2$CO$_3$, and oxalic acid were taken as precursors in their stoichiometric ratio. First, 11 mmol of oxalic acid and 2.2 mmol of V$_2$O$_5$ were dissolved in 30 mL of DI water taken in a Duran bottle and magnetically stirred at 70$^o$C for 30 mins. Then, 6.6 mmol of NaH$_2$PO$_4$ and 0.11 mmol of Na$_2$CO$_3$ are dissolved in 10 mL of DI water, magnetically stirred separately, and later drop wise added to the above solution. The Duran bottle was closed, heated, and magnetically stirred overnight at 80$^o$C. After that, the Duran bottle was opened, heated at 80$^o$C, and stirred until a gel was formed, which was later vacuum-heated overnight to remove any excess water.  The obtained dried precursors were grinded for 2 hrs and heated at 750$^o$C for 12 hrs in a high-temperature tube furnace under an Ar+H$_2$ (9:1) environment \cite{Ren_NE_16}. 

{\bf{Physical characterization:}}

We conducted x-ray diffraction measurements using a Panalytical Xpert 3 instrument at room temperature, employing CuK$_{\alpha}$ radiation with a wavelength of 1.5406 $\AA$. The Raman spectrum was collected using a Renishaw inVia confocal microscope equipped with a 532 nm wavelength laser and 2400 lines per mm grating. The elemental composition/mapping was checked through energy dispersive x-ray spectroscopy (EDX) using an Oxford-made LN2 free SDD X-max 80 EDS detector. The surface morphology was examined through Field Emission Scanning Electron Microscopy (FE-SEM) using a JEOL JSM-7800F Prime system. 

{\bf{Half-cell fabrication:}}

To conduct the electrochemical test, initially, half cells were prepared with NVP/C as the working electrode, sodium metal anode as the counter/reference electrode, and 1M NaPF$_6$ dissolved in ethylene carbonate (EC) and diethyl carbonate (DEC) in ratio of 1:1 as electrolyte. The electrode preparation process involved creating slurries of the active materials. Specifically, for NVP/C, a mixture of the active material, carbon (super P), and polyvinylidene fluoride (PVDF) in a ratio of 70:20:10 was utilized. N-methylpyrrolidone (NMP) was used as the solvent for the slurry. Following this, the slurry was applied onto a battery-grade aluminum foil using the doctor's blade casting technique. The coated foil was dried overnight at 80$^o$C to ensure complete solvent evaporation. After calendaring, 12~mm diameter electrodes were precisely cut out for further utilization. The assembly of 2032 coin cells were carried out inside a glove box (UniLab Pro SP from MBraun) filled with argon while maintaining a controlled atmosphere with O$_2$ and H$_2$O levels below 0.1 ppm. 

{\bf{Full-cell fabrication:}}

To construct the full cells, we use NVP/C as cathode and HC (purchased from Shaldong Gelon LIB Co., Ltd., having a median particle size (D50) of 9 $\mu$m and a specific surface area (SSA) of $\le$ 5 m$^2$/g) as an anode. The fabrication involves two steps: First, the HC anode was subjected to three cycles of sodiation in the half-cell configuration have sodium-metal as the counter/reference electrode. The slurry for the HC anode comprised a mixture of hard carbon, super-P, and PVDF in a mixture of 8:1:1 ratio. This slurry was coated onto a battery-grade copper current collector. These electrodes were then used to make half cells utilizing 1M NaPF$_6$ dissolved in EC:DEC as the electrolyte. The specific capacity of each cell was recorded, and the actual capacity of each cell was determined by multiplying it with the respective electrode weight of the cathode. The cells were disassembled in a fully charged state inside the glove box with an open-circuit voltage (OCV) of $\approx$1.1 V.

We use this sodiated hard carbon electrode material for the fabrication of full cell batteries. The weight of the electrodes was carefully selected to match the actual capacity of both the anode and cathode materials, respectively. These cells were assembled using the same electrolyte employed in the initial cycling of HC. The OCV of the NVP/C half cells was approximately 2.5 V, and the Hard-carbon half cells were disassembled at an OCV of roughly 1.1 V. Therefore, it is anticipated that the successfully fabricated full-cell will have an expected OCV in the vicinity of 1.4 V.

{\bf{Electrochemical characterization:}}

The electrochemical performance is tested through detailed analysis of electrochemical impedance spectroscopy (EIS) and cyclic voltammetry (CV), which were performed using the Biologic VMP-3 model. The EIS covered a frequency range from 10 mHz to 100 kHz, with an {\it ac} voltage amplitude of 10 mV applied at the cell's OCV state. The galvanostatic charge-discharge (GCD) profiles were obtained using a Neware battery analyzer (model no.: BTS400).

\begin{figure*} 
\includegraphics[width=7.2in]{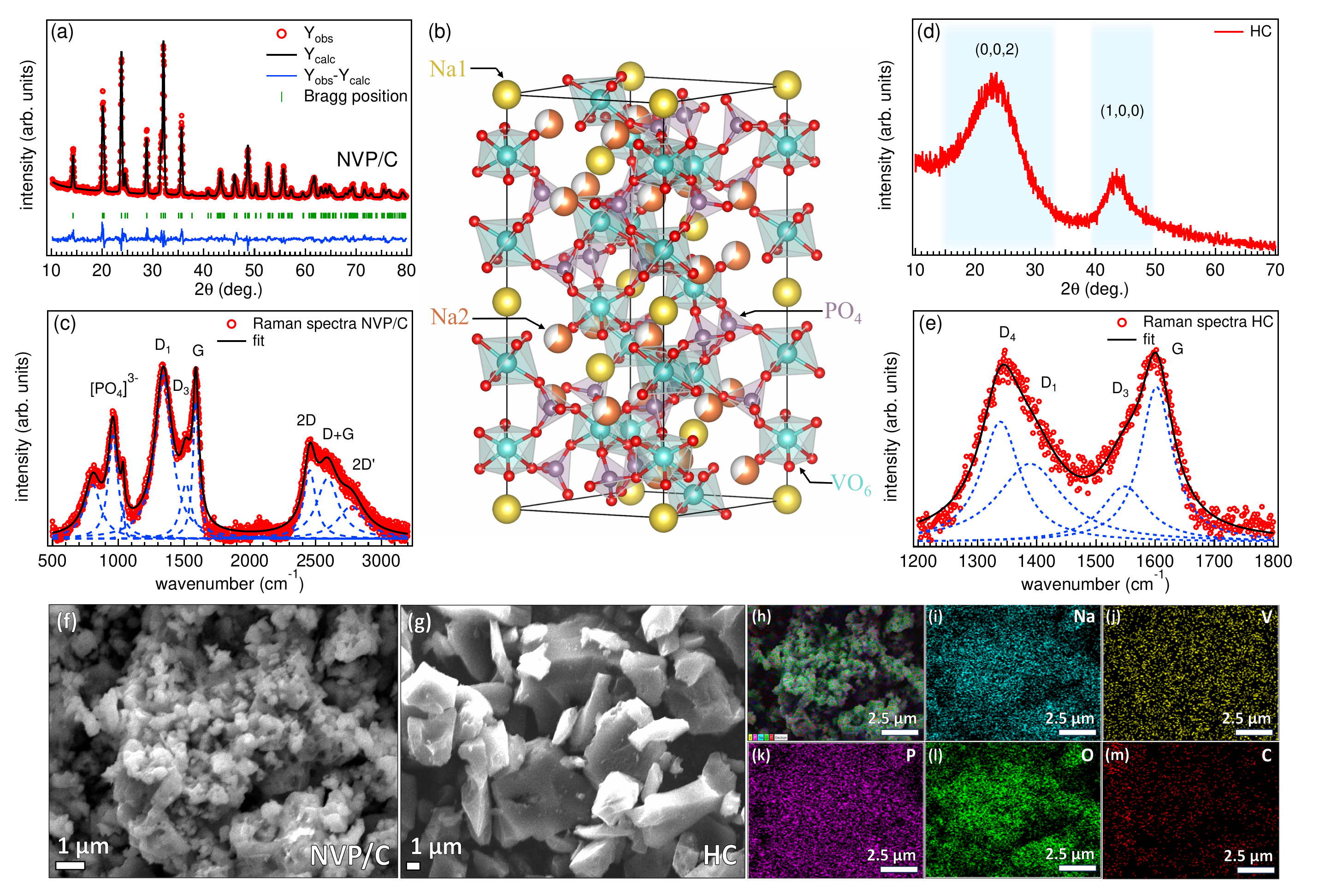}
\caption {(a) The Rietveld refined XRD pattern, (b) the crystal structure generated via VESTA software using refined parameters, and (c) the Raman spectrum of the prepared NVP/C sample. (d, e) The XRD pattern and the Raman spectrum of the HC. The FE-SEM images of (f) NVP/C and (g) HC samples. (h) The EDX layered image of NVP/C and the corresponding elemental mapping of (i) sodium (cyan), (j) vanadium (yellow), (k) phosphorus (pink), (l) oxygen (green), and (m) carbon (red).} 
\label{XRD}
\end{figure*}

\section{\noindent ~Results and discussion}

\begin{figure*}
\includegraphics[width=7.1in]{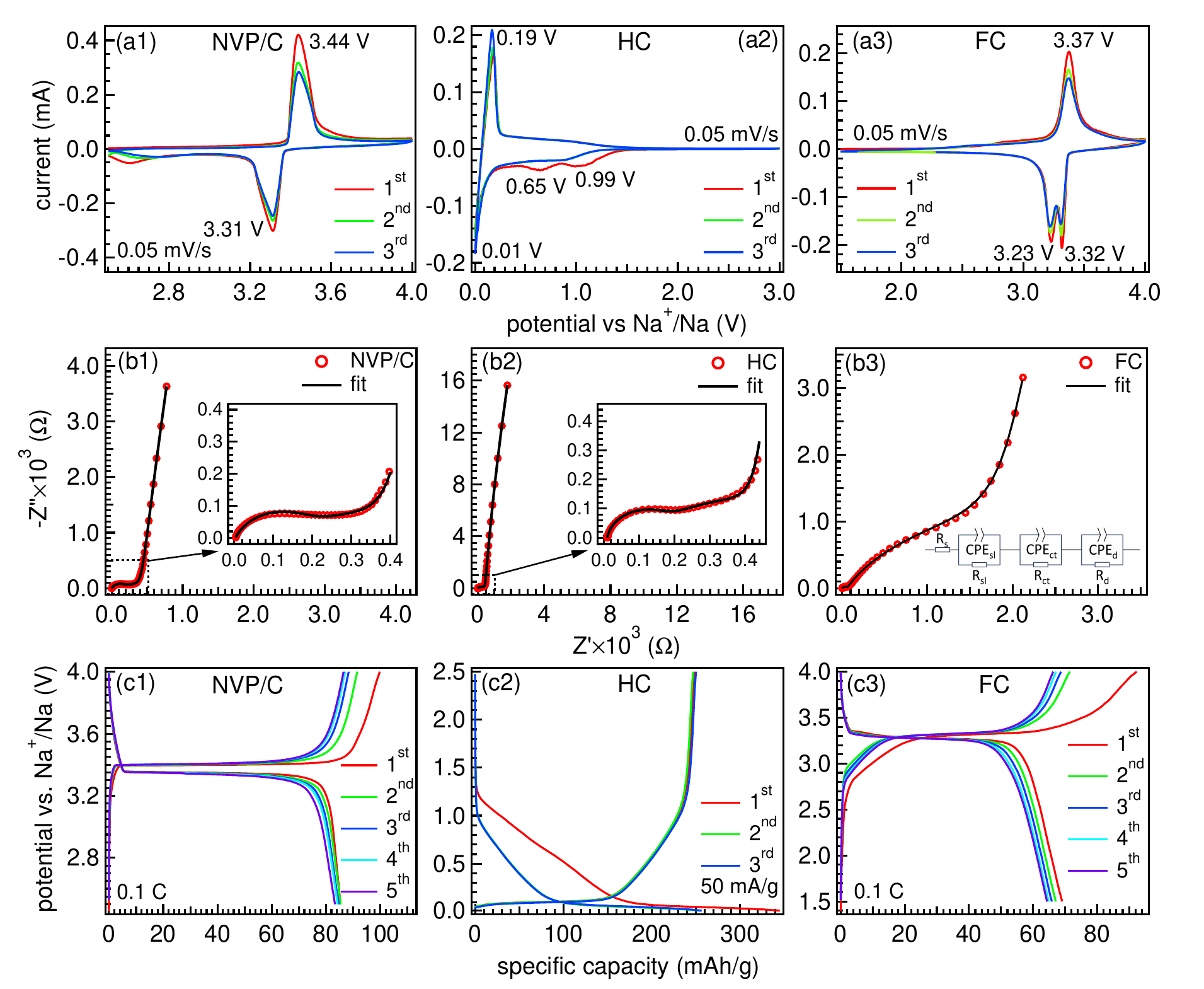}
\caption {First three cyclic voltammetry profiles at a scan rate of 0.05 mV/s in a potential range of (a1) 2.5--4.0 V for the NVP/C half cell, (a2) 0.01--3.0 V for the HC half cell, and (a3) 1.5--4.0 V for the HC$||$NVP/C full cell. The electrochemical impedance spectroscopy of the (b1) fabricated NVP/C half cell, (b2) the HC half cell, and (b3) the HC$||$NVP/C full cell with the corresponding equivalent circuit. The zoomed view is shown in the inset of (b1) and (b2). The galvanostatic charge-discharge profiles of (c1) NVP/C half cell in a potential range of 2.5--4.0 V at 0.1 C for 5 cycles, (c2) the HC half cell in a potential range of 0.01--2.5 V at 50 mA/g for 3 cycles, and (c3) the full cell in a potential window of 1.5--4.0 V at 0.1 C for 5 cycles.} 
\label{GCD}
\end{figure*}

Figure~\ref{XRD}(a) presents the XRD pattern along with the Rietveld refinement which confirms the reflections corresponding to a rhombohedral lattice and dimensions $a$($b$) = 8.72~\AA~ and $c$ = 21.78~\AA, classified under the space group R-3c, consistent with ref.~\cite{Saravanan_AEM_13}. The atomic occupancies and Wyckoff positions are provided in Table~I. 
\begin{table}[h]
		\label{Refine}
		\caption{Detailed structural information of NVP/C extracted from Rietveld refinement}
\begin{tabular}{p{1.1cm}p{1.3cm}p{1.3cm}p{1.3cm}p{1.1cm}p{1cm}}
\hline
\hline
 Atom&x&y&z&Occ.&Site\\
\hline
          Na1 &0.00000&0.00000 & 0.00000 & 1.000 & 6b \\ 
	       Na2 &0.63625&0.00000 & 0.25000 & 0.626 & 18e \\
                V &0.00000&0.00000 & 0.14728 & 1.000 & 12e \\
                 P &0.28492&0.0000 & 0.25000 & 1.000 & 18e \\
	 O1 &0.19422&0.16724 & 0.09202 & 1.000 & 36f \\
	 O2 &0.03414&0.21398 & 0.19213 & 1.000 & 36f \\
\hline
\hline
\end{tabular}
\end{table}
As depicted in Fig.~\ref{XRD}(b), the sodium ions (Na$^+$) are situated at the 6b (Na1) and 18e (Na2) positions, V$^{3+}$ ions fill the 12c octahedral sites, while P$^{5+}$ at the 18e tetrahedral sites \cite{Jian_AFM_14}. The Raman spectroscopy was employed to investigate the characteristics of the synthesized NVP/C, as illustrated in Fig.~\ref{XRD}(c). The pronounced D band observed at 1341.5 cm$^{-1}$ indicates the  defects, vacancies and edge structures of carbon. Concurrently, the G band, detected at 1589.5 cm$^{-1}$ is representative of the graphitic carbon. The intensity ratio between the G and D bands, I$_G$/I$_D$ is found to be 0.85, which underscores the disordered configuration of the carbon layer enveloping the NVP particles \cite{Xiong_AS_21}. Additionally, the three distinct signals detected at around 805 cm$^{-1}$ and 960 cm$^{-1}$ can be attributed to the bridging mode, where the 1035.5 cm$^{-1}$ belong to the stretching mode of the [PO$_4$] tetrahedra \cite{Jian_AEM_13, Dwivedi_AAEM_21}. The XRD profile of the HC exhibits two broad peaks at the 2$\theta$ value of approximately 23.5\degree and 44\degree, which are corresponding to the (002) and (100) crystal planes of disordered graphite domains. These features are indicative of the distinctive traits commonly observed in HC \cite{Xiao_AEM_18}. The Raman spectrum reveals distinctive features marked as D$_1$, D$_3$, D$_4$, and G bands, with peak positions at 1389 cm$^{-1}$, 1548.5 cm$^{-1}$, 1339 cm$^{-1}$, and 1602.5 cm$^{-1}$, respectively. The D$_1$ band, attributed to disorder-induced modes, signifies the disordered carbon structure in HC. The D$_3$ and D$_4$ bands are the indicators of sp$^3$ hybridized carbon, implying the existence of minute crystalline domains within the disorderly matrix. Additionally, the G band, aligned with the E$_{2g}$ vibrational mode in graphite, emphasizes the graphitic quality inherent in the carbon material \cite{Sadezky_C_05}. The microstructure of NVP/C and HC is examined through FE-SEM, as shown in Figs.~\ref{XRD}(f, g), respectively. The image reveals aggregated nanoparticles of NVP/C with a particle size of less than 1 $\mu$m. On the other hand, an irregular morphology having a particle size within the range of 5-10 $\mu$m is observed for HC. Additionally, Figs.~\ref{XRD}(h--m) present the layered EDX image and elemental mapping of Na, V, P, O, and C, providing evidence of the uniform distribution of these constituent elements in the sample. 

\begin{figure*}
\includegraphics[width=7.1in]{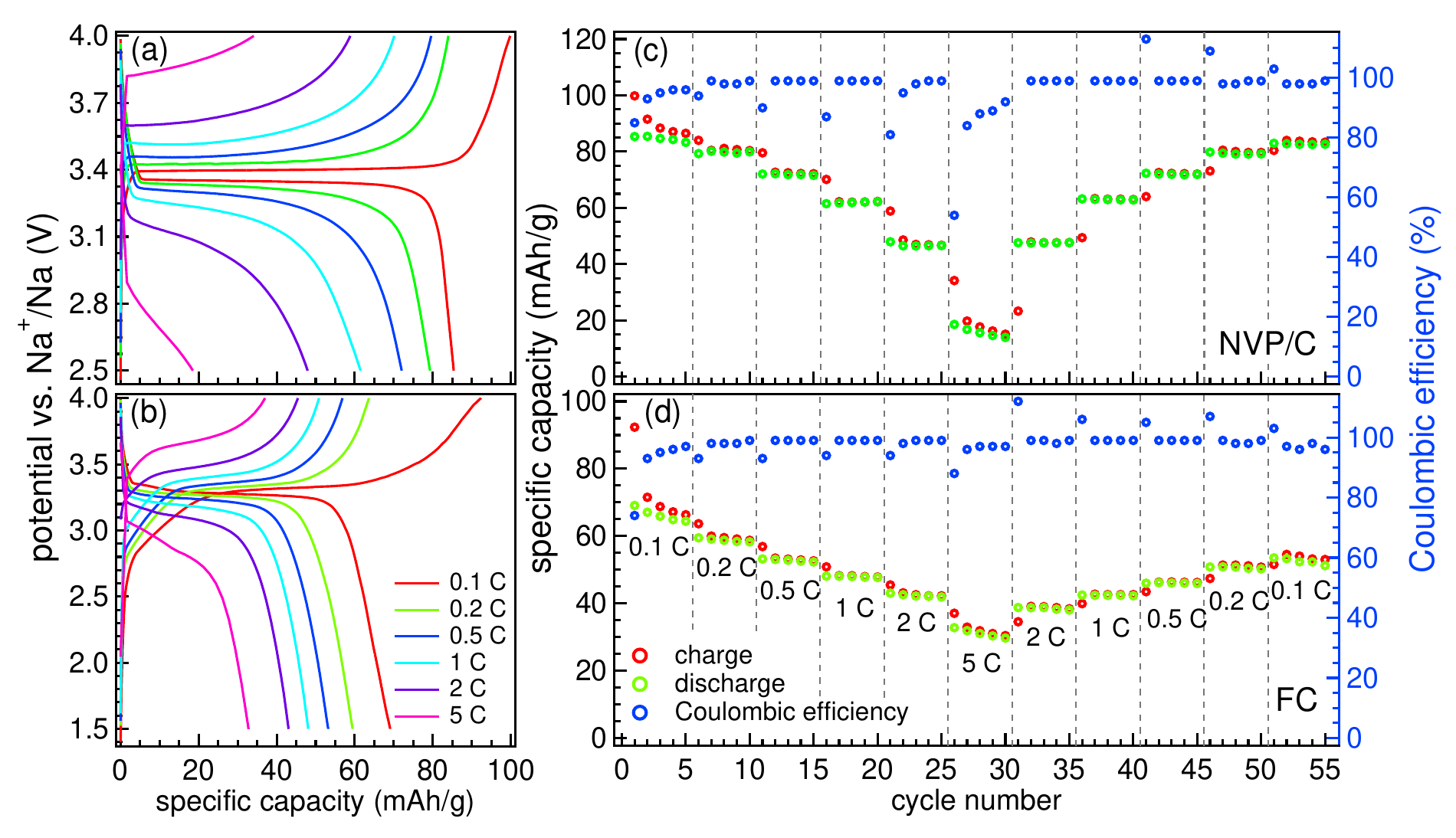}
\caption {The first cycle of the GCD profiles at each current rate from 0.1 C to 5.0 C for (a) the NVP/C half cell in the potential range of 2.5--4.0 V, and (b) the HC$||$NVP/C full cell in the potential range of 1.5--4.0 V. The corresponding rate capability, along with the Coulombic efficiency, are presented in (c) and (d) for both the cases, respectively.} 
\label{Rate}
\end{figure*}

Now, we first probe the electrochemical behavior of NVP/C and HC half cells. The cyclic voltammetry (CV) curves of NVP/C half cell are displayed in Fig.~\ref{GCD}(a1) up to three cycles at a scan rate of 0.05 mV/s within the potential window of 2.5-4.0 V against Na$^+$/Na. Notably, the oxidation peak at 3.44 V and the reduction peak at 3.31 V revealed the presence of the V$^{3+}$/V$^{4+}$ redox couple. While a slight decrease in peak current values post the first cycle indicated the formation of the solid electrolyte interface (SEI) layer, the peak voltage values remain consistent, with minimal polarization of around 0.1 V \cite{Sapra_WEE_21}. Likewise, the CV curves of the HC half cell are depicted in Fig.~\ref{GCD}(a2) covered the potential range of 0.01-3.0 V with a scan rate of 0.05 mV/s up to three cycles. In the first cathodic region, distinct peaks appeared at potential values of 0.01, 0.65, and 0.99 V. The anodic region displayed a sharp peak at 0.19 V and a broad region extending up to approximately 1.5 V, observed reversibly in both the anodic and cathodic regions (sharp peak at 0.01 V and a broad region extending to ~1.5 V). The irreversibility of the peaks at 0.65 and 0.99 V could be attributed to initial Coulombic losses due to the SEI formation \cite{Zhu_Ion_18}. The CV analysis of the full cell is conducted within the potential window of 1.5 to 4.0 V at a scan rate of 0.05 mV/s, as illustrated in Fig.~\ref{GCD}(a3). These CV profiles reveal a combination of features from the HC and NVP/C electrodes. In the anodic region, a distinct peak emerges at 3.37 V, accompanied by a preceding hump characteristic of the HC. Interestingly, the cathodic region displays two peaks at 3.32 and 3.23 V. This splitting of the cathode peak corresponds to the rearrangements in the local redox environment when two sodium ions sequentially insert into the Na(2) sites \cite{Li_JMCA_17, Feng_JMCA_17}. Notably, the polarization observed between the anodic and cathodic peaks is minimal around 0.05 V only. This value is lower than the potential gap observed in NVP/C half cell, indicating the superior electrochemical compatibility between the HC anode and the NVP/C cathode in full cell battery system. 

Further, the electrochemical impedance spectroscopy (EIS) measurements are conducted on freshly prepared cells (following a 9-hour rest) at open-circuit voltage (OCV). The obtained data are represented in the form of Nyquist plots, illustrated in Figs.~\ref{GCD}(b1--b3) for the NVP/C half cell, HC half cell, and HC$||$NVP/C full cell, respectively. These plots were fitted with an equivalent circuit displayed in the inset of Fig.~\ref{GCD}(b3). We find that the EIS behavior of NVP/C and HC half cells, in Figs.~\ref{GCD}(b1, b2), exhibit similar features: two depressed semicircles in the high and mid-frequency regions, indicative of solvation layer resistance (R$_{sl}$) and charge transfer resistance (R$_{ct}$), along with a straight line representing Na$^+$ ion diffusion in the bulk electrode at low frequencies. For the HC half cell, the R$_{sl}$ and R$_{ct}$ are measured to be 169.4 $\ohm$ and 261.3 $\ohm$, respectively, while for the NVP/C half cell, these values are found to be 165.4 $\ohm$ and 276.5 $\ohm$, respectively \cite{Thomas_JES_85, Gaddam_PCCP_21}. In contrast, the full cell exhibits a higher R$_{ct}$ value of 2241 $\ohm$, which is notably greater than the R$_{ct}$ values for both the NVP/C and HC half cells. This difference may be attributed to the blocking effect of the solid electrolyte interface (SEI) \cite{Pati_JMCA_22} on the HC anode, particularly because it was disassembled at a charged state and the same extracted material was used in the fabrication of the full cell battery. 

Moreover, the galvanostatic charge-discharge (GCD) measurements are performed for the NVP/C half cell at a current rate of 0.1 C (1 C = 117.6 mAh/g) within a potential window of 2.5--4.0 V, where the initial five cycles are depicted in Fig.~\ref{GCD}(c1). In these GCD profiles we find a flat plateau at around 3.4 V, which is attributed to the V$^{3+}$/V$^{4+}$ redox couple \cite{Li_JCIS_19}. The initial discharge capacity is observed around 85 mAh/g with the very good capacity retention in five cycles, reflecting an impressive stability rate of over 97\% with respect to the initial value. Also, the GCD profiles of the HC half cell, taken in a potential window of 0.01-2.5 V at a current density of 50 mA/g, revealed two distinct regions: sloping and flat, as shown in Fig.~\ref{GCD}(c2). The sloping region is indicative of Na$^+$ ion storage in defected carbon sites, while the flat region corresponds to intercalation of Na$^+$ ion between graphene layers and the filling of nano-pores \cite{Dou_MT_19, Bommier_NL_15}. The initial discharge capacity reached around 351 mAh/g, but upon charging, only 252 mAh/g was achieved, which is 71.2\% of the first discharge capacity due to the initial loss from irreversible surface reactions \cite{Stevens_JECS_00}. However, in the second cycle, the charging capacity almost matches the discharge capacity, having 96\% of the Coulombic efficiency. The GCD profiles of the full cell measured between 1.5 V and 4.0 V at 0.1 C current rate are presented in Fig.~\ref{GCD}(c3). Here, it is crucial to establish the voltage window with reference to the anode; therefore, the effective voltage values are obtained by subtracting the anode voltage from the cathode voltage, and we observe a combined effect of both the anode and cathode profiles. The GCD curves of the full cell featuring an ascending region starting around 2.7 V and stabilizing at 3.4 V, providing a working voltage of approximately 3.3 V, which found to be relatively higher than reported 2.7~V in ref.~\cite{Jian_AEM_13}. During charging, the Na$^+$ ions move from the cathode to the anode, and the ascending part reflects the filling of defect carbon sites, while the flat region signifies reaching the plateau for the HC anode. Interestingly, we find the stable charge-discharge capacity of 65--70 mAh/g (slight variation based on the mass loading of the cathode material) for the HC$||$NVP/C full cell up to five cycles, see Fig.~\ref{GCD}(c3). These values are found to be consistent with reported iparameters, as summarized in Table~II. 

\begin{table}[h]
\label{Compare}
\caption{Performance comparison of the sodium ion full cell with reported parameters.} 
\begin{tabular}{p{3.2cm}p{1.3cm}p{2.7cm}p{0.8cm}}
\hline
\hline
Anode$||$Cathode  & \begin{tabular}[c]{@{}l@{}}working \\ voltage\end{tabular} & specific capacity  & ref.  \\
\hline
NVP/C-T$||$NVP/C-T                                               & 1.7 V                                                      & 50.9 mAh/g at 0.4 C w.r.t. anode          & \cite{Liu_AMI_19}                  \\
MoTe$_2$$||$NVP                                                     & 2 V                                                        & 230 mAh/g at 0.2 A/g w.r.t. anode        & \cite{Panda_NE_19}                 \\
HC$||$NVP/C                                                      & 2.75 V                                                     & unsatisfactory w.r.t. cathode & \cite{Jian_AEM_13}                 \\
HCC$||$NVP                                                       & -                                                          & 46 mAh/g at 100 mA/g w.r.t cathode       & \cite{Akcay_AEM_21}                \\
GDHC$||$NVP                                                      & 2.4 V                                                      & 80 mAh/g at 0.05 C w.r.t cathode          & \cite{Vali_B_19}                   \\
\begin{tabular}[c]{@{}l@{}}NC@MoSe$_2$@rGO\\ $||$NVOPF\end{tabular} & 2.1 V                                                      & 286.7 mAh/g at 25 mA/g w.r.t. anode     & \cite{Roy_BS_21}                   \\
HC$||$NVP/C                                                      & 3.3 V                                                      & 70 mAh/g at 0.1 C w.r.t. cathode        & present work. \\

\hline
\hline
\end{tabular}
\end{table}

To test the rate capability, we further tested the electrochemical performance of the NVP/C half-cell and the HC$||$NVP/C full-cell battery. The first cycle galvanostatic charge-discharge (GCD) profiles are shown in Figs.~\ref{Rate}(a, b) for the NVP/C half-cell and HC$||$NVP/C full cell, respectively, at various current densities from 0.1 C to 5 C. The corresponding rate capability and the Coulombic efficiency at each current rate are presented in Figs.~\ref{Rate}(c, d). In the NVP/C half cell, we got a specific capacity of around 83 mAh/g at 0.1 C with remarkable capacity retention of 97.2\% after testing up to 55 cycles at various current rates. However, the specific capacity of around 70 mAh/g is observed in the full-cell at 0.1 C, which decreased to around 60 mAh/g having the retention of 77.5\% after 55 cycles at various current rates \cite{Cao_ASCE_20}. Nevertheless, in terms of high current rate performance, the full-cell outperformed the NVP/C half-cell at 5 C. The full cell retained 47.5\% of its initial capacity, while the NVP/C half-cell could hold only 21.7\% when the current rate increases from 0.1 C to 5 C. This suggests a complementary interaction between the HC anode and NVP/C cathode, consistent with the CV results.

\begin{figure}[h] 
\includegraphics[width=3.4in]{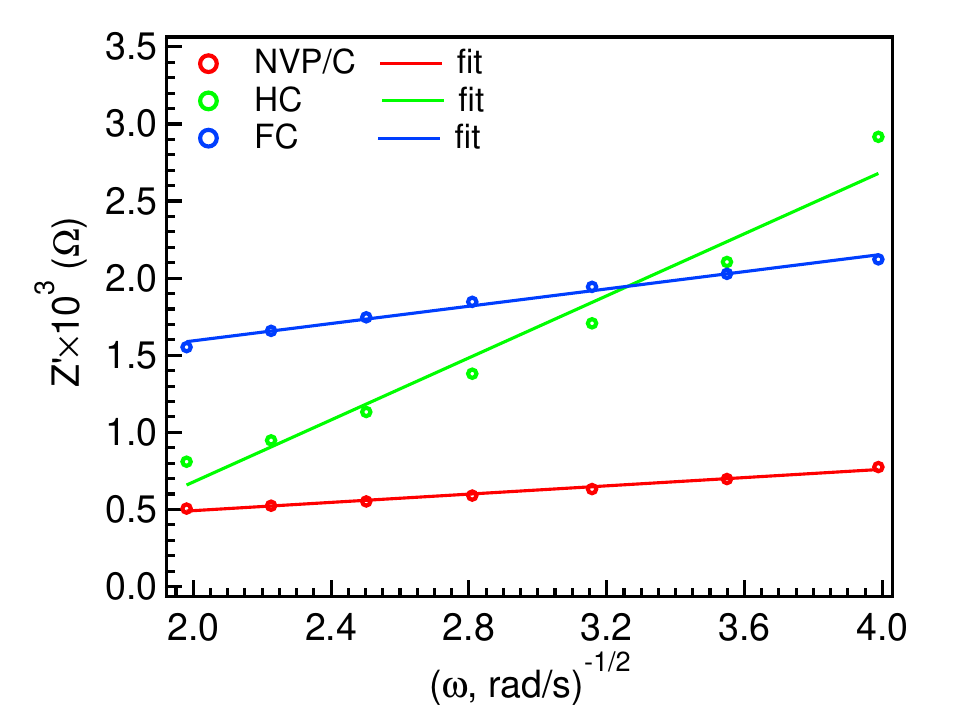}
\caption {The plot of the real part of impedance versus the inverse square root of the angular frequency in the Warburg region obtained from EIS data shown in Figs.~2(b1--b3)}
\label{Warburg}
\end{figure}

In order to understand the kinetics mechanism in these cells, we have done further analysis of the EIS data and estimated the diffusion coefficient using the following equations \cite{Chandra_EA_20}:-
\begin{equation}
D_{Na^+}=\frac{R^2 T^2}{2A^2\eta^4F^4C^2\sigma^2}
\label{diffusion}
\end{equation}
\begin{equation}
Z'=R_S +R_{CT}+\sigma\omega^{-0.5}
\label{warburg}
\end{equation}
where R, T, A, $\eta$, F, C, and $\sigma$ are gas constant, temperature, the geometrical surface area of the electrode, number of sodium ions participated in the reaction, the Faraday's constant, Na concentration in the electrode material (mol/cm$^3$) and the Warburg coefficient, respectively. To find the Warburg coefficient ($\sigma$), we plot the real part of impedance (Z$'$) as a function of the inverse square root of the angular frequency ($\omega^{-0.5}$). We then fitted these data points using equation~\ref{warburg}, see Fig.~\ref{Warburg}, where the estimated values of $\sigma$ are found to be 134.2, 1006.7 and 280.4 $\ohm/s^{1/2}$ for the NVP/C half cell, HC half cell and the full cell, respectively. The significantly higher value of $\sigma$ for the HC half cell suggests that the electrochemical process in the case of HC is primarily limited by diffusion, with mass transport occurring at a comparatively slower rate when compared with the NVP/C half cell as well as the full cell \cite{Lazanas_AMS_23}. By considering all the values in equation~\ref{diffusion}, the estimated values of the diffusion coefficient are: 0.28$\times$10$^{-12}$ cm$^2$/s for the NVP/C half cell, 0.27$\times$10$^{-13}$ cm$^2$/s for the HC half cell, and 0.54$\times$10$^{-13}$ cm$^2$/s for the full cell battery.  

\begin{figure}[h] 
\includegraphics[width=3.4in]{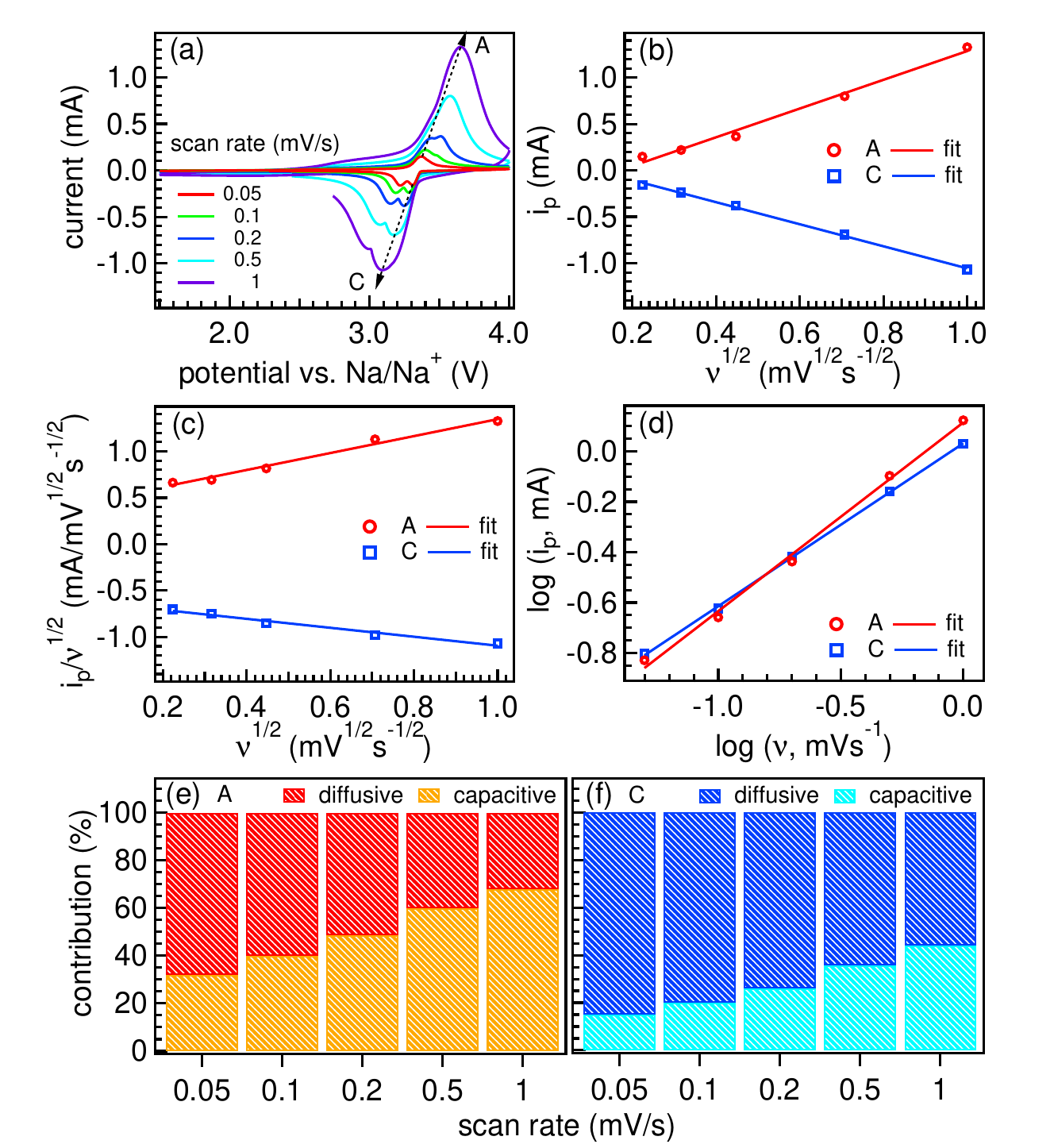}
\caption {(a) The CV curves of the HC$||$NVP/C full cell at different scan rates in the voltage range of 1.5--4.0 V, (b) the linear fit of the peak current i$_p$ of A and C as a function of the square root of the scan rate, (c) the linear fit between i$_p$/$\upsilon$$^{1/2}$ as a function of $\upsilon$$^{1/2}$, (d) the linear fit between log(i$_p$) as a function of log($\upsilon$), the capacitive and diffusive contributions for the peaks (e) A and (f) C at different scan rates.}
\label{CV_FC}
\end{figure}

The cyclic voltammetry (CV) measurements were done on the fabricated full cell (having NVP/C as cathode and HC as anode) at different scan rates ranging from 0.05 to 1 mV/s in a voltage window of 1.5 to 4.0 V. Interestingly, the CV curve displayed prominent peaks in the cathodic as well as in the anodic region as marked by C and A in Fig.~\ref{CV_FC}(a), respectively. We observe that the peak height increases from 0.1 mA to 1.3 mA when the scan rate vary from 0.05 mV/s to 1 mV/s. To probe the kinetic behavior of the sodium-ion battery, we calculate the diffusion coefficient values for both cathodic and anodic peaks using the Randles-Sevcik equation \cite{Saroha_ACSO_19}.
\begin{equation}
i_p = (2.69 \times 10^5) A D^{\frac{1}{2}} C \eta^{\frac{3}{2}} \upsilon^{\frac{1}{2}}
\end{equation}
where $i_p$ (mA) is the peak current, A (cm$^2$) is the electrode surface area, D (cm$^2$/s) is the diffusion coefficient, C (mol/cm$^3$) is the concentration of Na ions in bulk, $\eta$ is the number of transferred electrons in the redox reaction, and $\upsilon$ (mV/s) is the applied scan rate. The peak current obtained from the oxidation/reduction peaks is plotted as a function of the square root of the scan rate $\upsilon$, shown in Fig.~\ref{CV_FC}(b), and the values of $D$ are calculated using the slope we got from the linear fitting of the $i_p$ versus $\upsilon^{\frac{1}{2}}$. The diffusion coefficient values are found to be 8$\times$10$^{-11}$ and 4.7$\times$10$^{-11}$ cm$^2$/s for A, and C peaks, respectively. These values match well with Na$^+$ ion diffusion in both NVP/C and HC half cells, justifying the coordinated performance of the HC$||$NVP/C full-cell \cite{Song_JMCA_14, Ledwoch_EA_22, Ohishi_APCA_22}. 

Finally, to attribute the contribution of diffusion and surface-controlled reactions, we further analyzed the data using the Dunn method \cite{Manish_CEJ_23, Chen_NatCom_15}.
\begin{equation}
\frac{i}{\upsilon^{\frac{1}{2}}} = k_1 \upsilon^{\frac{1}{2}} + k_2
\end{equation}
where k1 and k2 are the coefficients for the respective capacitive and diffusion currents, which are calculated by linear fitting of the $i$/$\upsilon^{1⁄2}$ versus $\upsilon^{1⁄2}$, as shown in Fig.~\ref{CV_FC}(c), the values for k1 and k2 are found to be 0.91 and 0.43 for anodic peak and 0.48 and 0.61 for the cathodic peak, respectively. These values indicate that the diffusion part is dominant in cathodic and has a high capacitive contribution in the case of anodic reaction. In order to determine the contribution quantitatively at different scan rates, we show the log($i$) versus log($\upsilon$) plot in Fig.~\ref{CV_FC}(d) and performed the linear fit  using the power law \cite{Wang_JPCC_07}:
\begin{equation}
i = a \upsilon^b
\end{equation}
here $i$ is the current, $\upsilon$ is the scan rate, and $a$ and $b$ are the fitting parameters. The value of $b$ determines the nature of the sodium ion charge/discharge reaction mechanism. If the $b=$ 0.5, the storage process is diffusion controlled, and if the  $b=$ 1, the process is capacitive, and the intermediate value of $b$ marks the contribution from both. The obtained values of $b$ are 0.75 and 0.65 for anodic and cathodic peaks, respectively. These contributions are shown at various scan rates in Figs.~\ref{CV_FC}(e, f) for the anodic and cathodic peaks, respectively. These values confirm the dominating diffusive controlled process at low scan rates and the capacitive contribution found to increase at higher scan rates. For example, the diffusive contribution in the total current at 0.05 mV/s is around 68 and 85\% for anodic and cathodic peaks, respectively, which drops to 32 and 56\% at a high scan rate of 1 mV/s. For both the anodic and cathodic reactions, the general trend follows the lowering of diffusion-controlled contribution with increasing the scan rates. \\

\section{\noindent ~Conclusions}

In summary, we synthesized NVP/C cathode material and investigated its electrochemical performance in a full cell configuration with HC as an anode. The phase of the material was confirmed using Rietveld refinement of the XRD pattern and Raman spectroscopy measurement. In the half-cell configuration, both the NVP/C and HC demonstrated good electrochemical performance against the sodium metal anode, yielding a reversible discharge capacity of 85 mAh/g at 0.1 C and a charge capacity of 252 mAh/g at 50 mA/g, respectively. The HC$||$NVP/C full cell demonstrated a decent specific capacity of approximately 70 mAh/g at 0.1 C and exhibited a high-rate capability, with a capacity of around 35 mAh/g at 5 C. The diffusion coefficient for the HC$||$NVP/C full cell is found to be 5--8$\times$10$^{-11}$ cm$^2$/s for both the anodic and the cathodic peaks using thorough cyclic voltammetry (CV) analysis. The collaborative effect of HC and NVP/C enabled a high operational voltage of approximately 3.3 V, with a minimal polarization of only 0.05 V. This investigation into the HC$||$NVP/C full cell contributes valuable insights into the development of cost-effective high energy density sodium-ion battery system for energy storage applications, potentially advancing the affordability of these devices for the common public. 

\section{\noindent ~Acknowledgments}
Madhav thanks CSIR-HRDG for the fellowship support and Central Research Facility (CRF) of IIT Delhi for the FE-SEM measurements. We acknowledge Department of Science and Technology (DST), Govt. of India for financially supporting the research facilities for sodium-ion battery project through {\textquotedblleft}DST-IIT Delhi Energy Storage Platform on Batteries" (project no. DST/TMD/MECSP/2K17/07), and from SERB-DST through a core research grant (file no.: CRG/2020/003436). We thank IIT Delhi for providing research facilities for sample characterization (XRD and Raman at the physics department).

\end{document}